\documentclass[%
reprint,
 amsmath,amssymb,
 aps,
]{revtex4-2}

\usepackage{graphicx}
\usepackage{dcolumn}
\usepackage{bm}
\usepackage{hyperref}
\hypersetup{colorlinks=true,
    linkcolor=blue,
    filecolor=magenta,      
    urlcolor=cyan,
    }
\usepackage{chemfig}
\usepackage{soul}
\usepackage[version=4]{mhchem}
\usepackage{subcaption}
\usepackage{graphicx}
\newcommand{\ket}[1]{\left| {#1} \right\rangle}
\usepackage{float}
\UseRawInputEncoding

\begin{document}

\preprint{APS/123-QED}

\title{H\"uckel Molecular Orbital Theory on a Quantum Computer: \\ A Scalable System-Agnostic Variational Implementation with Compact Encoding}
\author{Harshdeep Singh}
\email{harshdeeps@kgpian.iitkgp.ac.in}
\affiliation{%
 Center of Computational and Data Sciences, Indian Institute of Technology, Kharagpur, India
}%
\author{Sonjoy Majumder}%
\email{sonjoym@phy.iitkgp.ac.in}
\affiliation{%
  Department of Physics, Indian Institute of Technology, Kharagpur, India}%
\author{Sabyashachi Mishra}%
\email{mishra@chem.iitkgp.ac.in}
\affiliation{%
  Department of Chemistry, Indian Institute of Technology, Kharagpur, India}%

\date{\today}

\begin{abstract}
H\"uckel molecular orbital (HMO) theory provides a semi-empirical treatment of the electronic structure in conjugated $\pi$-electronic systems. A scalable system-agnostic execution of HMO theory on a quantum computer is reported here based on a variational quantum deflation (VQD) algorithm for excited state quantum simulation. A compact encoding scheme is proposed here that provides an exponential advantage over the direct mapping and allows quantum simulation of the HMO model for systems with up to $2^N$ conjugated centers in $N$ qubits. The transformation of the H\"uckel Hamiltonian to qubit space is achieved by two different strategies: a machine-learning-assisted transformation and the Frobenius-inner-product-based transformation. These methods are tested on a series of linear, cyclic, and hetero-nuclear conjugated $\pi$-electronic systems. The molecular orbital energy levels and wavefunctions from the quantum simulation are in excellent agreement with the exact classical results. The higher excited states of large systems, however, are found to suffer from error accumulation in the VQD simulation. This is mitigated by formulating a variant of VQD that exploits the symmetry of the Hamiltonian. This strategy has been successfully demonstrated for the quantum simulation of \ce{C_60} fullerene containing 680 Pauli strings encoded on six qubits. The methods developed in this work are system-agnostic and hence are easily adaptable to similar problems of different complexity in other fields of research.  

\end{abstract}

\maketitle


\section{Introduction}
\label{sec:intro}
\setcounter{equation}{0}
Quantum computing harnesses the principles of quantum mechanics to process information in fundamentally different ways than classical computers~\cite{q2, q1}. Unlike the classical bits, quantum bits or qubits can exist in a superposition of states and interact with each other via entanglement. Quantum computing exhibits significant potential for transforming different domains of science and technology~\cite{Advantage1, farhi2019quantum, jaramillo_2016_quantum, first}, but its capacity for catalyzing substantial change is notably evident in the discipline of chemistry~\cite{qintro1, qintro3, qintro6, cao_2019_quantum}. 
Despite the major advances in the classical simulation of quantum chemical systems~\cite{intro1, intro2}, classical computing encounters challenges in effectively simulating the inherent quantum nature of electrons in a molecule, thus limiting its accuracy in predicting and simulating molecular interactions with a high degree of precision~\cite{challenge2, challenge3}. In contrast, quantum computers operating with principles of quantum mechanics are particularly well-suited for simulating molecular quantum phenomena in the fields of catalysis, materials research, and drug discovery~\cite{drug1, drug2, drug3}. 

Despite the potential for revolutionizing computing, the current generation of quantum computers, the so-called Noisy Intermediate-Scale Quantum (NISQ) computers, are susceptible to external noise.
The constraints on the scale and duration of calculations arise from the limited number of qubits and the relatively short coherence duration. In the NISQ era of quantum computing, hybrid classical-quantum algorithms like Variational Quantum Eigensovlers (VQE) have emerged as the most effective tools for molecular electronic structure calculations in small systems like \ce{H_2, LiH, BeH_2}, and \ce{H_2O}~\cite{vqa, qintro4, qintro5}. These algorithms have been experimentally tested on different quantum computing architectures like photonic processors~\cite{photonic} and trapped-ion processors~\cite{trapion}. The use of unitary coupled cluster methods~\cite{qintro2, qintro7}, and the advances in error-resistant algorithms, and quantum error correction~\cite{error1, error2, error3, halder_2023_development} add further to the utility of these variational algorithms.

\begin{figure*}[!]
\centering
\includegraphics[width=\textwidth]{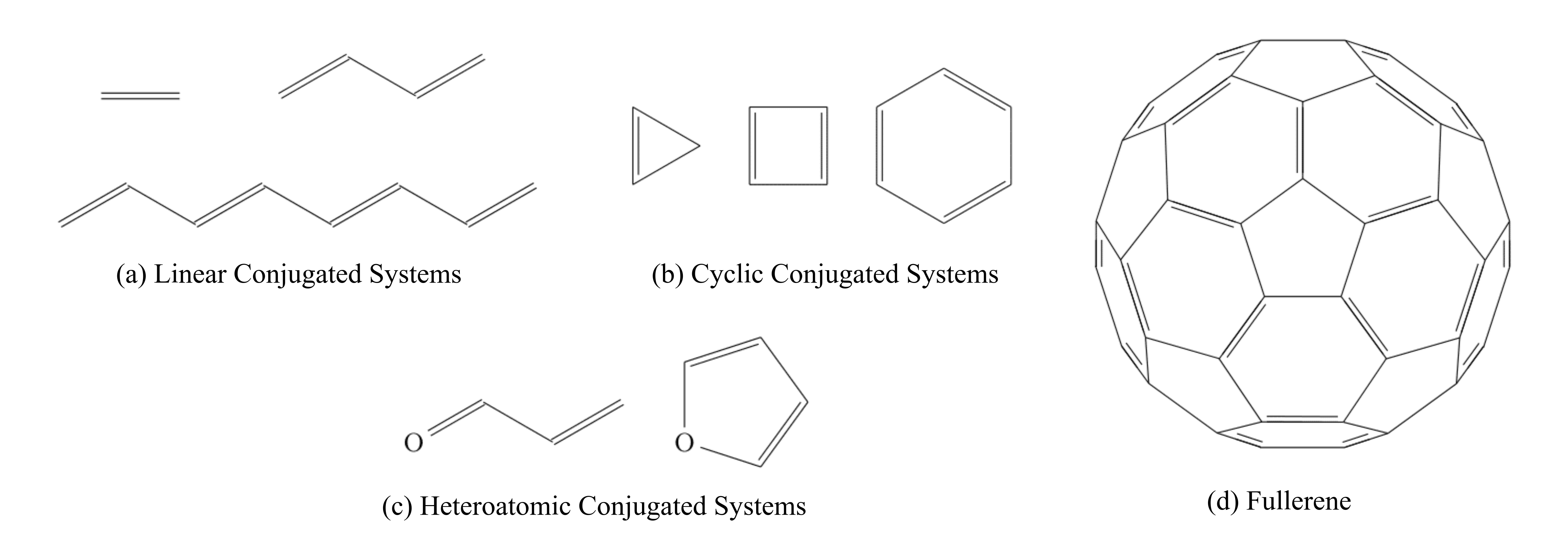}
\caption{Different molecules simulated under HMO theory on a quantum system.}
\label{fig:molecules}
\end{figure*}

The current work focuses on executing the well-known semi-empirical H\"uckel Molecular Orbital (HMO) theory~\cite{https://doi.org/10.1002/jcc.20470} on a quantum computer by introducing a compact encoding scheme. The only quantum computation effort on the HMO theory available in the literature is by Yoshida et al. \cite{huckel}. However, they could compute only small molecules (up to Benzene) due to their direct encoding method, which required $n-$qubits for a $n-$carbon conjugated system. The compact encoding scheme employed in our work allows us to compute energy and wavefunctions for 2$^n$ conjugated carbon systems with an $n$-qubit system, an exponential improvement over the previous method in terms of qubits requirement. We have implemented a robust generalized method of transforming the HMO Hamiltonian matrix into the qubit space that enables efficient quantum simulation of HMO Hamiltonian for both homo-and hetero-nuclear conjugated systems with different architecture, such as linear, cyclic, or even poly-cyclic conjugated systems. We demonstrate the potentialities of the current implementation of quantum computing on a set of representative conjugated systems with varying size and character, including Fullerene \ce{C_{60}} (Figure~\ref{fig:molecules}).

\section{Theoretical Background}
\subsection{H\"uckel Molecular Orbital Theory}
\label{sec:huckel}
The HMO theory, formulated on the principle of $\sigma-\pi$ separation in planar conjugated systems, has remained a fundamental tool in theoretical chemistry for close to a century~\cite{Huckel1931-qi}, aiding in the understanding of electronic configuration, aromaticity, molecular stability, reactivity, and spectroscopy of planar conjugated organic molecules~\cite{huckel2, huckel3}. 
Within HMO theory, the electronic Schr\"odinger equation of a planar conjugated system is solved with a linear variational approach where the molecular wavefunction is expressed as a linear combination of the $p_z$ atomic orbitals ($\phi$) of $n$ carbon atoms, 
\begin{equation}
    \left| \psi \right \rangle = \sum_{i=1}^{n}c_i \ket{\phi_i}
\end{equation}
where, $c_i$s are the coefficients of the $p_z$ orbitals of individual carbon atoms, whose values are determined by minimizing the variational energy. The coefficients that minimize the energy are obtained from the non-trivial solution of the $n$ secular equations, represented in the following determinant form,
\begin{equation}
    \begin{vmatrix}
H_{11} - ES_{11} & \dots & H_{1n} - ES_{1n}\\
\vdots & \ddots & \vdots \\
H_{n1} - ES_{n1}  & \dots & H_{nn} - ES_{nn}
\end{vmatrix} \cdot \begin{vmatrix}
c_1\\
\vdots \\ 
c_n
\end{vmatrix} =0
\label{huck5}
\end{equation}
where, the energy integrals ($H_{ij}$) and the overlap integrals ($S_{ij}$) are defined as,
\begin{equation}
    H_{ij} = \langle \phi_i|{H}|\phi_j\rangle
\end{equation}
and
\begin{equation}
    S_{ij} = \langle \phi_i|\phi_j\rangle,
\end{equation}
respectively.
The diagonal terms in the Hamiltonian integrals ($H_{ii}$) are the so-called Coulomb integrals, while the non-diagonal terms ($H_{ij}$) are known as the resonance integrals. Instead of an explicit evaluation of the energy and overlap integrals, the HMO theory makes some empirical approximations, e.g., $S_{ij}=\delta_{ij}$, $H_{ii}=\alpha$, and $H_{ij}=\beta$ if $i$ and $j$ carbon atoms are directly bonded, else $H_{ij}=0$. 
The solutions of Equation~\ref{huck5} are conventionally expressed through the quantities $\alpha$ and $\beta$, which are obtained via some empirical rules~\cite{hetero1, hetero2}.
For example, the H\"uckel matrix of ethylene takes the form
\begin{equation}
    {H}_{\ce{C_2H_2}} = \alpha+ \beta \begin{bmatrix} 
0  & 1\\
1 & 0 
\end{bmatrix},
\end{equation}
whose eigenvalues represent the MO energy levels, $ E = \alpha \mp \beta$ and the corresponding eigenvectors indicate the molecular orbital composition. For homo-nuclear conjugated systems, $\alpha$ appearing at the diagonal elements acts as an additive constant, and $\beta$ is the only free parameter that determines the energy spectrum. 

For hetero-nuclear conjugated systems, the Coulomb and resonance integrals involving the heteroatoms (O, S, or N) are different from those involving only the carbon atoms. Hence, the Coulomb integrals for heteronuclear conjugated systems are modified as, $\alpha^{\prime} = \alpha + h \beta$.
Similarly, the resonance integrals in heteronuclear systems are modified as 
$\beta^{\prime} = k \beta$, where, $h$ and $k$ are system specific empirical parameters~\cite{hetero1, hetero2}.
For example, in the case of acrolein (\ce{C_3H_4O}), the correction constants are set as $h=1$ for the oxygen atom, $k=1.1$ for \ce{C=C}, and $k=0.9$ for \ce{C-C}, resulting in the final H\"uckel Hamiltonian matrix,
\begin{equation}
    {H}_{\ce{C_3H_4O}} = \alpha + \beta \begin{bmatrix}
0 & 1.1 & 0 & 0\\
1.1 & 0 & 0.9  & 0\\
0 & 0.9 & 0  & 1.1 \\
0 & 0 & 1.1 & 1
\end{bmatrix},
\end{equation}
with $\alpha$ as an additive constant, and $\beta$, as a multiplicative constant. 

The quantum computation of the HMO requires a transformation of the H\"uckel Hamiltonian to a qubit form (described in section~\ref{sec:transformation}) and a strategy to obtain the MO eigenvalues and eigenfunctions on quantum hardware, discussed next.

\subsection{Variational Quantum Eigensolver}
\label{sec:vqe}
VQE is a hybrid classical-quantum algorithm described as a parameterized quantum circuit that is trained with the help of a classical optimizer~\cite{vqa}. 
In a typical variational quantum algorithm (VQA), the first step is the preparation of the initial state, often taken as the default configuration of the qubits, i.e., $\displaystyle |\Psi_{\rm in}\rangle = |0 \cdots 0\rangle$.  The variational parameters ($\Vec{\theta}^{(i)}$) are then introduced into the circuit via the \textit{ansatz} that comprises various unitary rotation and entanglement gates ($U(\Vec{\theta}^{(i)})$). For a given set of variational parameters, the state of the system is given as $\displaystyle |\Psi(\Vec{\theta}^{(i)})\rangle = U(\Vec{\theta}^{(i)}) |\Psi_{\rm in}\rangle$. The energy (or any other observable) of the system in the given state is obtained by a measurement of the corresponding operator in qubit representation, which acts as the \textit{cost function} of the algorithm that needs to be minimized. With the Hamiltonian in qubit representation expressed in terms of Pauli strings $P_\alpha$ as 
\begin{equation}
    H_{\rm qubit} = \sum_\alpha a_\alpha P_\alpha,
\end{equation} 
the energy measurement of the state $\Psi(\theta_i)$ yields
\begin{equation}
        E(\Vec{\theta}^{(i)}) = \left \langle \Psi(\Vec{\theta}^{(i)}) \left| \sum_\alpha a_\alpha P_\alpha \right|\Psi(\Vec{\theta}^{(i)})\right \rangle.
        \label{eq:1}
\end{equation}
The variational parameters ($\Vec{\theta}^{(i)}$) are further trained by using a suitable classical optimizer to obtain an improved set of variational parameters ($\Vec{\theta}^{(i+1)}$) iteratively until convergence. The \textit{optimizer} constitutes the classical component, whereas the remaining part constitutes the quantum component of the hybrid VQE algorithm.

\subsection{Variational Quantum Deflation (VQD) Algorithm  for Excited States}
\label{sec:vqd}
While VQE offers the solution for the ground state of a problem, the solution of the excited states can be obtained with some modification to the VQE algorithm. One such algorithm employed in the current work is the variational quantum deflation (VQD) method~\cite{vqd}. While there are a few other algorithms available to find the excited state energies~\cite{excited1, excited2, excited3}, the VQD algorithm offers a solution for excited states using the same amount of qubits as the VQE algorithm, with a circuit depth that is at most twice as VQE implementation. The VQD algorithm exhibits resilience against control errors, possesses compatibility with error-mitigation schemes, and can be feasibly executed on the NISQ devices. Starting from the optimized VQE solution (Section~\ref{sec:vqe}), the VQD method finds the $k^{th}$-excited state of the system by optimizing the parameters $\theta_k$ for the state $|\Psi_k\rangle$ such that the function $F(\Vec{\theta}_k)$ is optimized, where
\begin{eqnarray}
      F(\Vec{\theta}_k) &=& \langle \Psi(\Vec{\theta}_k)|H_{\rm qubit}|\Psi(\Vec{\theta}_k)\rangle + \sum_{j=0}^{k-1} \gamma_j |\langle \Psi(\Vec{\theta}_k)|\Psi(\Vec{\theta}_j) \rangle|^2
\nonumber  \\ 
        &=&
        E(\Vec{\theta}_k) + \sum_{j=0}^{k-1} \gamma_j |\langle \Psi(\Vec{\theta}_k)|\Psi(\Vec{\theta}_j) \rangle|^2.
        \label{eq:2}         
\end{eqnarray}
where, $\{\Vec{\theta}_k\}$ is the optimized parameters of the $k^{th}$ energy state. This can be seen as optimizing $E(\Vec{\theta}_k)$ with an additional constraint that the current state $|\Psi(\theta_k)\rangle$ is orthogonal to the previous states $|\Psi(\Vec{\theta}_0)\rangle, |\Psi(\Vec{\theta}_1)\rangle, \cdots, |\Psi(\Vec{\theta}_{k-1})\rangle$. Here, $\gamma$ balances the contribution of each overlap term to the cost function and is generally computed as the mean square sum of the coefficients of the observable. This is equivalent to finding the ground state energy of a modified Hamiltonian at a stage $k$,
\begin{equation}
    H_k = H + \sum_{i=0}^{k-1} \gamma_i |i\rangle \langle i|
\label{eq:vqd_h}
\end{equation}
where, $|i\rangle$ is the $i^{th}$ eigenstate of the Hamiltonian $H$ with energy $E_i = \langle i|H|i\rangle $.

\begin{figure*}[!]
\centering
\includegraphics[width=\textwidth]{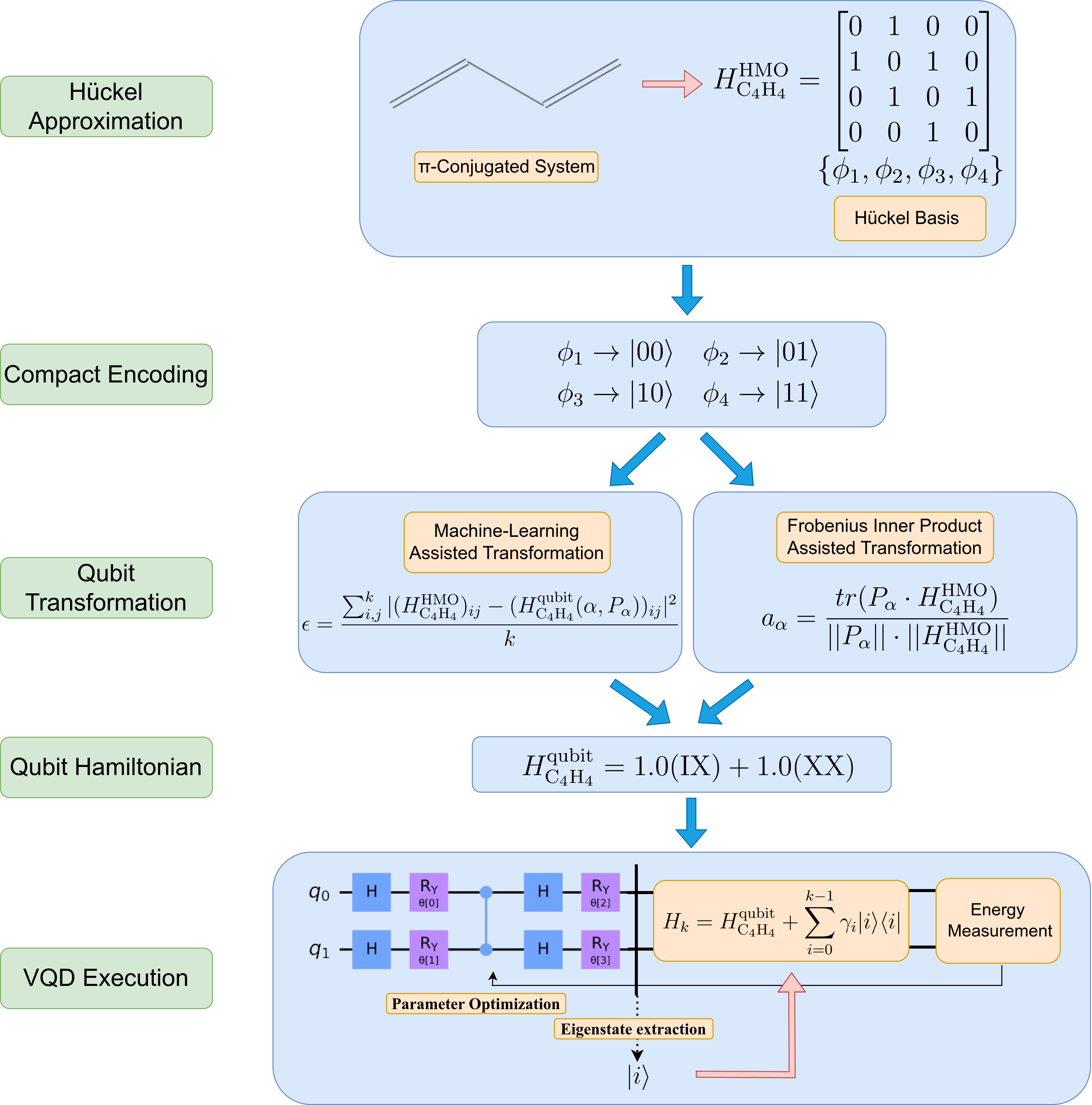}
    \caption{Schematic representation of the employed workflow for quantum simulation of HMO theory with \ce{C_4H_4} as a case study.}
\label{fig:resource}
\end{figure*}

\begin{table*}
\caption{\label{tab:table1} The HMO Hamiltonian (as a summation of Pauli strings) and the number of qubits required for VQD simulation of different conjugated molecules with the compact encoding scheme. Both machine-learning-assisted transformation and Frobenius-inner-product-based transformation yield identical qubit Hamiltonians.}
\begin{ruledtabular}
    \begin{tabular}{c|ccc}

Molecule  &  Hamiltonian & \# Qubits  \\
  & &   \\
\hline  
\hline  
C$_2$H$_4$  & $1.0 (\rm X)$ & 1 \\
\hline 
C$_3$H$_4$  & $0.5(\rm IX) + 0.5(XI) + 0.5(XX) + 0.5(XZ) + 0.5(YY) + 0.5(ZX)$ & 2 \\
\hline 
C$_4$H$_4$ & $1.0(\rm IX) + 1.0(XX)$ & 2  \\
\hline 
C$_4$H$_6$ & $1.0(\rm IX) + 0.5(XX) + 0.5(YY)$ & 2  \\
\hline 
C$_3$H$_4$O  & $0.25(\rm II) + 1.1(IX) - 0.25(IZ) + 0.45(XX) + 0.45(YY) - 0.25(ZI) + 0.25(ZZ)$& 2  \\
\hline 
C$_4$H$_4$O &  $0.25(\rm III ) + 0.425(\rm IIX ) + 0.25(IIZ) + 0.275(IXX ) + 0.275(IYY ) + 0.25(IZI) $ & 3 \\
   & $+ 0.25(\rm IZZ) - 0.025(IZX) + 0.20(XII) + 0.20(XIZ) + 0.275(XXX)  - 0.275(XYY)$   \\
  & $ + 0.20(\rm XZI) + 0.20(XZZ) + 0.275(YXY) + 0.275(YYX) + 0.25(ZII) + 0.425(ZIX)$   \\
& $ + 0.25(\rm ZIZ) + 0.275(ZXX) + 0.275(ZYY) + 0.25(ZZI) - 0.025(ZZX) + 0.25(ZZZ)$ \\ 
\hline 
C$_6$H$_6$ &  $0.75(\rm IIX) + 0.25(IXX) + 0.25(IYY) + 0.25(IZX) + 0.25(XIX) + 0.25(XXX) $ & 3  \\
   & $ - 0.25(\rm XYY) + 0.25(XZX) - 0.25(YIY) + 0.25(YXY) + 0.25(YYX) $ &  \\
  & $ - 0.25(\rm YZY) + 0.25(ZIX) + 0.25(ZXX) + 0.25(ZYY) - 0.25(ZZX)$ &  \\
\hline 
C$_8$H$_{10}$ &  $1.0(\rm IIX) + 0.5(IXX) + 0.5(IYY) + 0.25(XXX) - 0.25(XYY) $& 3 \\
   & $   0.25(\rm YXY) + 0.25(YYX) $ &  &

\label{table4}
\end{tabular}
\end{ruledtabular}
\end{table*}

\section{Methodology}
\subsection{Encoding Scheme}\label{sec:transformation}
The first step in executing H\"uckel MO theory on a quantum computer is to transform the problem from the fermionic to the qubit space. Recently, Yoshida et al~\cite{huckel} implemented a direct encoding technique, where each basis function ($\phi_i$), i.e., each $\pi-$electron is encoded into a qubit. Therefore, for \ce{C_2H_4}, the atomic orbital basis set $\{\phi_1, \phi_2\}$ is directly mapped as,
\begin{equation}
    \phi_1 \rightarrow |01\rangle \quad {\rm and} \quad \phi_2 \rightarrow |10\rangle.
\end{equation}
In the direct encoding scheme, the qubit requirement grows linearly $\mathcal{O}(N)$ with the number of C atoms in the conjugated system. This scheme results in a large number of 'spurious' states. For example, in \ce{C_2H_4}, the states $|00\rangle$ and $|11\rangle$ do not constitute any physical state. The number of such states increases exponentially with the system size, making the encoding highly susceptible to noise. In this work, we introduce a compact encoding scheme where each basis function can be encoded into individual states of the qubit system, i.e.,
\begin{equation}
    \phi_1 \rightarrow |0\rangle \quad {\rm and} \quad \phi_2 \rightarrow |1\rangle.
\end{equation}
With our compact encoding scheme the system size that can be simulated with $N$ qubits scales as $\mathcal{O}(2^N)$, thus providing an exponential advantage over the direct mapping.

\subsection{H\"uckel matrix in qubit representation}
The H\"uckel Hamiltonian described in section~\ref{sec:huckel} needs to be expressed in terms of a linear combination of Pauli strings. In the direct mapping scheme employed by Yoshida et al, the  H\"uckel Hamiltonian in the qubit space takes a simple form, 
\begin{equation}
    H = \frac{\beta}{2}\sum_{(m,n)\in\{bonds\}}({\rm X}^{m}{\rm X}^{n} + {\rm Y}^{m}{\rm Y}^n).
\label{eq:hamil}
\end{equation}
However, the applicability of this direct encoding scheme is limited to simple linear or monocyclic homo-atomic conjugated systems. With the ability of the compact encoding scheme to deal with larger systems, we now require a generalized scalable transformation scheme to treat conjugated systems of arbitrary complexity. 

In the most general form, the H\"uckel Hamiltonian of a  conjugated molecular system with up to $2^N$ conjugated centers encoded with $N$ qubits can be expressed as a sum of $4^N$ Pauli strings each of length $N$, i.e., 
\begin{equation}
    H_{\rm qubit} = a_1 ({\rm II\cdots I}) + a_2 ({\rm II\cdots X}) + \cdots +     a_{4^{N}} ({\rm ZZ\cdots Z}), \label{eq:qubit_H}
\end{equation}
where $a_i$ are the coefficients that determine the contribution of the concerned Pauli string to the H\"uckel Hamiltonian. The coefficients can be determined via (i) machine-learning-assisted scheme or (ii) the Frobenius inner-product-assisted scheme. 

\subsubsection{Machine Learning Assisted Encoding}
The machine-learning-assisted encoding is more of a brute-force technique of finding out the coefficients of different Pauli strings that represent the original Hamiltonian. The training is done by defining an error $\epsilon$, which in our case was chosen as the mean squared difference between the elements of the original H\"uckel Hamiltonian in $\pi$-electron basis $(H_{ij})$ and the predicted H\"uckel Hamiltonian in qubit space ($H^{\rm pred}_{ij}$), i.e.,
\begin{equation}
    \epsilon = \frac{\sum_{i,j}^{k} |H_{ij} - H^{\rm pred}_{ij} |^2}{k}.
    \label{eq:error}
\end{equation}
In the case of butadiene (\ce{C_4H_4}), the number of qubits required is $2$, and the complete set of 16 Pauli strings are \{II, IX, IY, IZ, XI, YI, ZI, XY, YX, YZ, ZY, XZ, ZX, XX, YY, ZZ\}. The qubit Hamiltonian in the general form appears as, 
\begin{equation}
    H^{\rm qubit}_{\ce{C_4H_4}} = a_1({\rm II}) + a_2({\rm IX}) + \cdots + a_{16}({\rm ZZ}).   
\end{equation}
After the training process, 
the qubit Hamiltonian for \ce{C_4H_4} that minimizes the error (Equation~\ref{eq:error}) is obtained as,
\begin{equation}
    {H}^{\rm qubit}_{\ce{C_4H_4}} = \mathrm{1.0(IX) + 1.0(XX)}.
\label{eq:c4h4}
\end{equation}

\subsubsection{Frobenius Inner Product Assisted Encoding}
The Frobenius inner product of two matrices $K$ and $L$ of the same dimensions ($m \times n$) is defined as the sum of the element-wise products of the matrices~\cite{frob}, i.e.,
\begin{equation}
    \langle K, L \rangle_F = \sum^{m}_{i}\sum^{n}_{j} K^*_{ij} \cdot L_{ij} = tr(K^{\dagger}L).
\end{equation}
The value of the Frobenius inner product represents the degree of similarity or correlation between the matrices.
If the matrices exhibit orthogonality, their Frobenius inner product will assume a value of zero, signifying their independence. On the other hand, a greater inner product implies a greater degree of similarity between the matrices. 

If a matrix $K$ can be written as a summation of $n$ matrices $L_i$'s that form a complete set, i.e., $K=\sum^n_i a_i L_i$, the coefficients $a_i$ can be obtained from the Frobenius inner product of $K$ and $L_i$ 
as
\begin{equation}
    a_i = \frac{\langle K, L_i \rangle_F}{{||K||\cdot||L_i||}} .
\end{equation}
where, $||K||$ is the Frobenius Norm of the matrix $K$, defined as,
\begin{equation}
    ||K|| = \sqrt{tr(K^{\dagger}K)}.
\end{equation}
In the present context, we use the Frobenius inner products to evaluate the coefficients of the $4^N$ Pauli strings (each of length $N$) required for constructing the qubit Hamiltonian of dimension $2^N \times 2^N$  in a $N$-qubit space (Equation~\ref{eq:qubit_H}). 
The coefficient for the Pauli string $P_\alpha$ is given by, 
\begin{equation}
    a_{\alpha} = \frac{tr(P_\alpha \cdot H)}{{||P_\alpha||\cdot||H||}}.
\label{eq:alpha}
\end{equation}
Since $P_\alpha$ and $H$ are unitary matrices,  $||P_\alpha||\cdot||H|| =2^N $  for an $N$-qubit problem. For example, in the case of \ce{C_4H_4}, among the 16 possible Pauli strings (of length 2 each), the non-zero Frobenius inner products are obtained only for the Pauli strings IX and XX, with 
\begin{equation}
    a_{\rm IX} = \frac{tr({\rm IX} \cdot {H}^{\rm qubit}_{\ce{C_4H_4}})}{2^2} = 1.0
    \label{eq:c4h4matrix}
\end{equation}
\begin{equation}
    a_{\rm XX} = \frac{tr({\rm XX}\cdot {H}^{\rm qubit}_{\ce{C_4H_4}})}{2^2} = 1.0.
\end{equation}
This results in the same Hamiltonian as in Equation~\ref{eq:c4h4}.
Since the size of the predicted Hamiltonian $(2^N \times 2^N)$ is dictated by the number of qubits $(N)$, but the size of the molecular system can be any matrix of size $(M \times M)$, where $M$ is the number of $\pi-$electrons in the system, an appropriate number of dummy layers in the form of rows and columns made up of zero elements can be introduced to the molecular Hamiltonian without loss of generality to make sure molecules of different sizes can be resolved by the algorithm. For example, the H\"uckel matrix for the \ce{C_3H_4} molecule for a  2-qubit transformation can be written as,
\begin{equation}
    {H}_{\ce{C_3H_4}} = \begin{bmatrix}
0 & 1 & 1 & 0\\
1 & 0 & 1  & 0\\
1 & 1 & 0  & 0 \\
0 & 0 & 0 & 0
\end{bmatrix}.
\end{equation}

\subsection{Quantum Simulation Settings}
Using the H\"uckel Hamiltonian in qubit space, we used the VQD algorithm to find the energy spectra and eigenfunctions of the considered molecule. All the simulations were carried out in IBM's Qiskit~\cite{qiskit} framework. We employed four different settings, namely, \textit{exact} (a numerical diagonalization of the H\"uckel matrix with a classical computer), \textit{ideal} (VQD simulation with an ideal quantum circuit simulator), \textit{noisy} (VQD simulation with a noisy quantum circuit simulator), and \textit{realistic} (VQD simulation with a noisy quantum circuit simulator embedded with external noise from IBM Cairo device). Limited-memory Broyden-
Fletcher-Goldfarb-Shanno bound (L\_BFGS\_B) optimizer was used for the ideal quantum settings and Simultaneous Perturbation Stochastic Approximation (SPSA) was employed for the noisy settings. Since the method employed is variational, the simulations were carried out over one hundred times to find the average result of the energy spectrum of different molecules in different simulation settings.

\begin{table*}
\caption{\label{tab:table2} The HMO energy levels ($E_n$ in the units of resonance integral $\beta$) from direct diagonalization of the HMO Hamiltonian (Exact) and VQD simulation with ideal quantum circuit simulator (Ideal), with noisy quantum circuit simulator (Noisy),  and with external (IBM Cairo) noise embedded quantum circuit simulator (Realistic).}
\begin{ruledtabular}
    \begin{tabular}{c|cccccccccc}

Molecule & Simulation Setting & $E_1$ & $E_2$ & $E_3$ & $E_4$ & $E_5$ & $E_6$ & $E_7$ & $E_8$ \\
\hline  
\hline  
C$_2$H$_4$ & Exact & -1.0 & 1.0 &  &   &  &  & \\
             & Ideal & -1.0 & 1.0 &  &   &  &  & \\
               & Noisy & -0.9646 & 0.9770 &  &  &  &  & \\
               & Realistic & -1.0937 & 0.8729 &  &  &  &  & \\
\hline  
C$_3$H$_4$ & Exact & -2.0 & 1.0 & 1.0 &  &  &  & \\
             & Ideal & -2.0 & 1.0 & 1.0 &  &  &  & \\
               & Noisy & -1.9914 & 0.9448 & 1.0106 &  &  &  & \\
               & Realistic& -2.1868 & 0.7900 & 0.8452 &  &  &  & \\
\hline  
C$_4$H$_4$ & Exact &-2.0 & 0 & 0 & 2.0 &  &  & \\
            & Ideal &-2.0 & 0 & 0 & 2.0 &  &  & \\
               & Noisy & -1.9891 & -0.0205 & -0.0178 & 1.9994 &  &  & \\
              & Realistic& -1.9022 & -0.0325 & 0.0088 & 1.7023 &  &  & \\
\hline  
C$_4$H$_6$ & Exact & -1.6180 & -0.6180 & 0.6180 & 1.6180 &  &  & \\			
         & Ideal & -1.6180 & -0.6180 & 0.6180 & 1.6180 &  &  & \\	
            & Noisy & -1.7721 & -0.6183 & 0.5212 & 1.6125 &  &  & \\				
            & Realistic & -1.5518 & -0.6616 & 0.3602 & 1.3522 &  &  &  \\
\hline  								

C$_3$H$_4$O & Exact & -1.9275 & -1.0738 & 0.4583 &  1.5430 & & & \\
             & Ideal & -1.9275 & -1.0738 & 0.4583 &  1.5430 & & &\\
               & Noisy & -1.9329 & -1.0719 & 0.4713  &  1.5217 & & &\\
                & Realistic & -1.7739 & -1.0180 & 0.4134  &  1.3419 & & &\\
\hline  
C$_4$H$_4$O & Exact & -2.6524 & -1.2908 & -0.7384 & 1.0432 & 1.6384 & &\\
             & Ideal & -2.6524 & -1.2908 & -0.7384 & 1.0432 & 1.6384 & &\\
               & Noisy & -2.6955 & -1.3438 & -0.6676  & 0.9973 & 1.6575 & &\\
                & Realistic & -2.3955 & -1.1720 & -0.5812  & 0.6938 & 1.1900 & &\\
\hline

C$_6$H$_6$ & Exact & -2.0 & -1.0 & -1.0 & 1.0 & 1.0 & 2.0 &   \\		
         & Ideal & -2.0 & -1.0 & -1.0 & 1.0 & 1.0 & 2.0 &   \\	
         & Noisy & -1.9697 & -1.01855 & -0.9043 & 0.8903 & 0.9704 & 1.9692 &   \\		
         & Realistic & -1.5673 & -0.6906 & -0.6241 & 0.5968 & 0.6726 & 1.5595 &   \\		

\hline  								
C$_8$H$_{10}$ & Exact & -1.8793 & -1.5320 & -0.9999 & -0.3472 & 0.3472 & 0.9999 & 1.5320 & 1.8793\\
        & Ideal & -1.8794 & -1.5356 & -1.0160 & -0.3604 & 0.3404 & 1.0 & 1.5321 & 1.8794\\
         & Noisy &  -1.8814 & -1.3151 & -0.8393 & -0.3031 & 0.3416 & 0.9325 & 1.3750 & 1.5626 \\
         & Realistic & -1.4353 & -1.0991 & -0.6367 & -0.1874 & 0.2970 & 0.7739 & 0.9425 & 1.2847

\label{table4}
\end{tabular}
\end{ruledtabular}
\end{table*}

\section{Results and Discussion}
\subsection{Compact Encoding Scheme}
During the execution of quantum computing operations, several types of errors arise, including but not limited to decoherence errors, read-out errors, and crosstalk errors~\cite{errors}. These problems are inherent to the noisy intermediate-scale quantum (NISQ) devices. In order to mitigate the impact of noise, it is necessary to impose limitations on the overall quantity of qubits employed in computations, ensuring that the noise magnitude remains within a manageable range for error mitigation purposes. 

In the present work, we employ a compact encoding scheme that can use $N$ qubits to simulate HMO for a system up to 2$^N$ conjugated centers (Figure~\ref{fig:resource}). This encoding scheme provides an exponential advantage over the direct mapping employed by Yoshida et al~\cite{huckel}. For example, using three qubits the compact encoding can simulate \ce{C_8H_10} (see Table~\ref{tab:table1}), while the same system would require 8 qubits with direct mapping. An immediate outcome of the compact encoding scheme is the significant reduction in the number of spurious states for systems with conjugated centers not equal to $2^N$. For any system with $2^N$ conjugated centers, the compact encoding results in no spurious states, while for the other systems, there will be some states that do not represent any physical state and are generally discarded. The comparison of the spurious states from direct and compact encoding shows that the number of spurious states remains reasonably small with the compact encoding scheme, making it resource-efficient; see Figure~\ref{fig:resource} for a comparison. A system like \ce{C_60} fullerene would require 60 qubits with $2^{60}-60$ spurious states with direct encoding, while the compact encoding scheme requires six qubits with merely four spurious states.

\begin{figure}[!]
\centering
\includegraphics[width=0.45\textwidth]{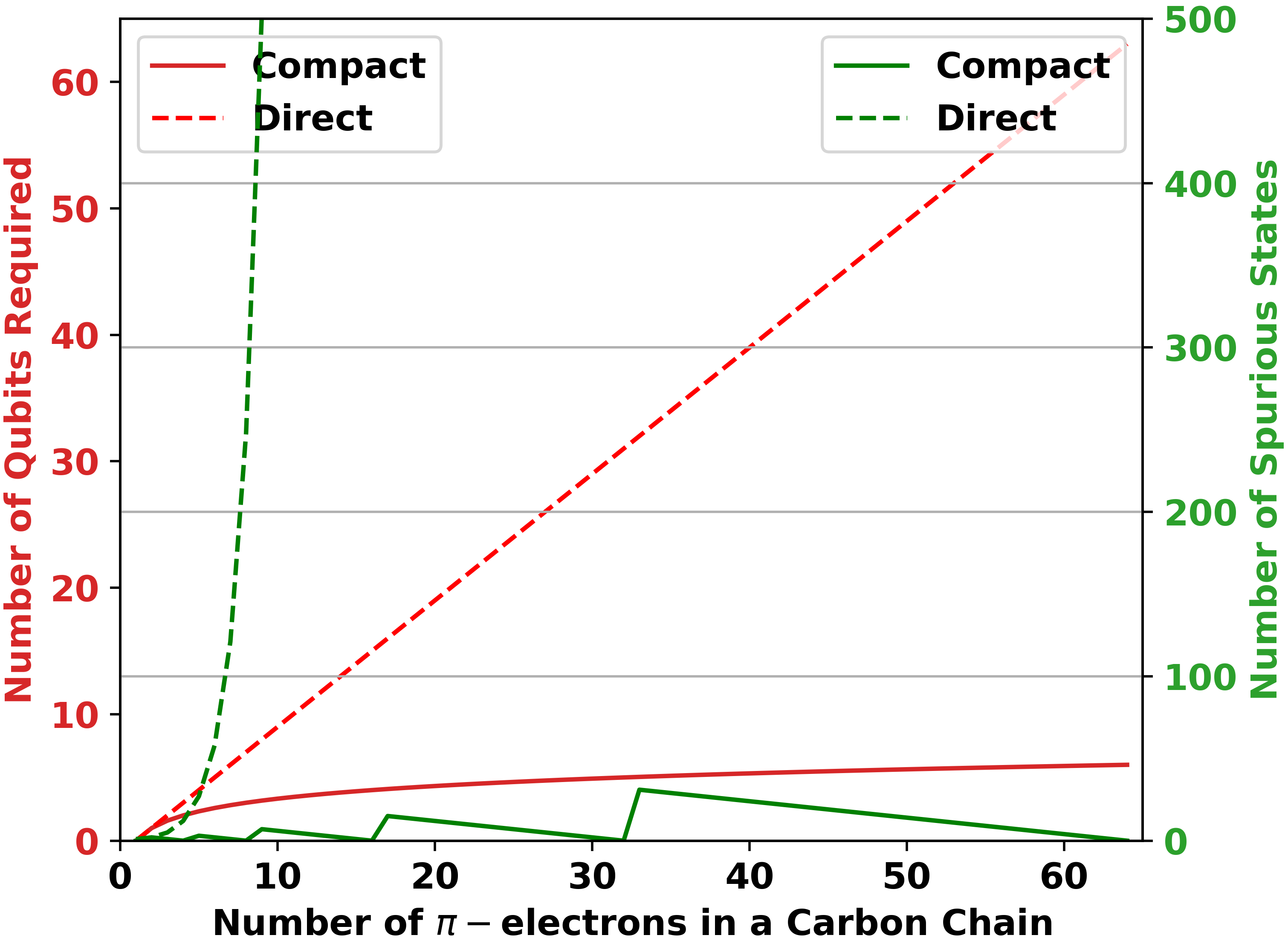}
    \caption{Resource requirement for executing HMO theory on a quantum device with the compact and direct encoding scheme, highlighting the number of qubits required and the number of spurious states for a linear carbon chain consisting of up to 64 $\pi-$electrons. }
\label{fig:resource}
\end{figure}


The scalability of the compact encoding scheme brings forth a new challenge, i.e, an efficient transformation of large H\"uckel matrices to qubit space. 
To that end, our first approach, namely, the machine-learning-assisted transformation, offers a generic and robust transformation method that is system-agnostic. Such a method does not have any prerequisite for the shape, size, and nature of the matrix, nor does it require a completeness of the basis set. 
The HMO qubit Hamiltonians for different conjugated systems are rather simple for smaller systems, while they get quite complicated as system size increases (Table~\ref{tab:table1}). The brute-force nature of the machine-learning method gets extremely expensive for larger systems. By design, this method neither offers a guarantee of convergence nor a unique solution. For a system with 6 qubits, where the number of trainable parameters becomes $4^6$, the machine learning method requires more than $10^6$ iterations for convergence, which is computationally quite expensive.

This limitation is mitigated, to a large extent, by the Frobenius inner-product method of Hamiltonian transformation. The latter offers a straightforward and systematic way of evaluating the contributions from each of the possible Pauli strings to the Hamiltonian matrix. The transformation in the case of a 6-qubit problem could be readily achieved by this method, which involves a large number of matrix multiplications. The convergence to a unique solution is guaranteed by this method as long as the employed basis forms a complete set. Of course, this step also becomes exponentially expensive for larger systems as it involves inner-product evaluation of the Hamiltonian with every possible element ($4^N$) of the Pauli basis. Nonetheless, conjugated systems of sizes that are of general interest to the chemists can be handled by either of the two proposed transformation schemes.
 
\subsection{Choice of Optimizer}
\begin{figure}[!]
\centering
\includegraphics[width=0.5\textwidth]{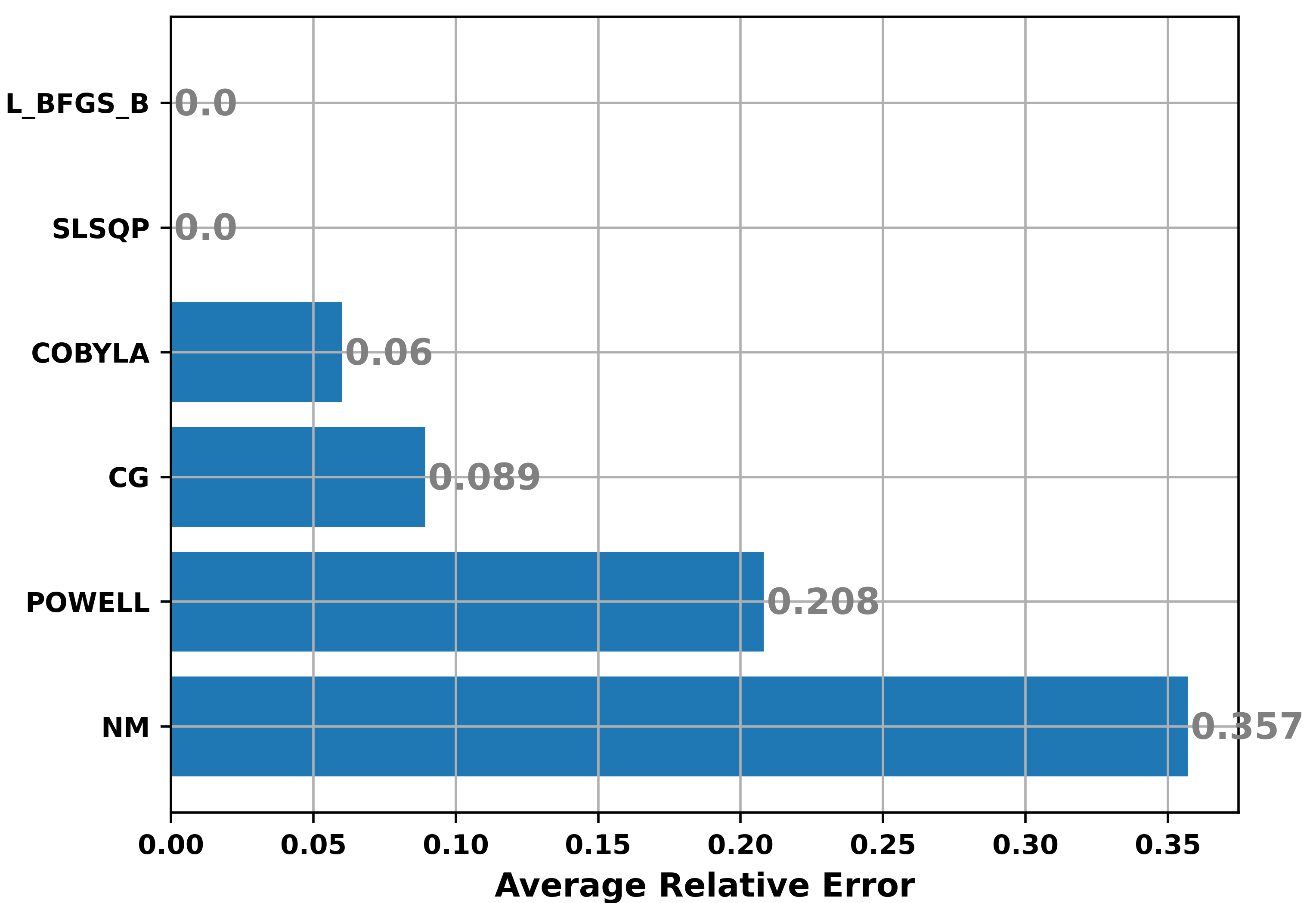}
    \caption{Performance of different classical optimizers (labelled in the ordinate) for the simulation of \ce{C_{16}H_{10}} in ideal quantum settings. 
    The average errors $\displaystyle\left(\frac{1}{n}\bigg |\sum^{n}_{k=1}\frac{|E_k^{\rm exact}| - |E_k^{\rm VQD}|}{|E_k^{\rm exact}|}\bigg |\right)$, of all 16 eigenvalues are indicated in the figure.}
\label{fig:optimizer}
\end{figure}

One of the major influencing factors in the performance of the VQD algorithm is the choice of classical optimizers. Our earlier benchmarking study for VQE algorithms in quantum chemistry applications~\cite{hd}, has shown that the conjugate-gradient (CG) optimizer and SPSA optimizer are the best-performing optimizers in the ideal and noisy quantum settings, respectively. Hence, all the simulations were carried out with these optimizers initially until we encountered a  noticeable performance lag in the CG optimizer. While it performed extremely well for the smaller molecules, it started to get highly erroneous for the larger molecules, evident from a high overall error, as highlighted in~Figure~\ref{fig:optimizer}. 
This can be attributed to the fact that the CG optimizer requires the matrix of the system to remain positive-definite, which is not the case for HMO matrices~\cite{cg}. In the context of the VQD algorithm, since one keeps adding the contributions of previous eigenstates into the Hamiltonian matrix (Equation~\ref{eq:vqd_h}), the performance of the optimizer deteriorates as system size increases.

This prompted further investigation for the right optimizer for the VQD algorithm. To that end, different commonly used classical optimizers were tested across different molecules (Figure~\ref{fig:optimizer}). For smaller molecules, most of the optimizers perform reasonably well. The challenge appears in the simulation of larger molecules, such as \ce{C_8H_{10}} and \ce{C_{16}H_{10}}. Based on the average results, the L\_BFGS\_B, and sequential least squares programming (SLSQP) optimizers turn out to be the best-performing optimizers. All the results presented in this paper are obtained using L\_BFGS\_B optimizer in the ideal quantum setting. In noisy conditions, however, the noise-resistant SPSA remains the best choice, as the performance of all other optimizers is severely affected by the noise~\cite{hd}.

\subsection{Circuit Depth of the Employed Ansatz }
\begin{figure}[!]
\centering
\includegraphics[width=0.5\textwidth]{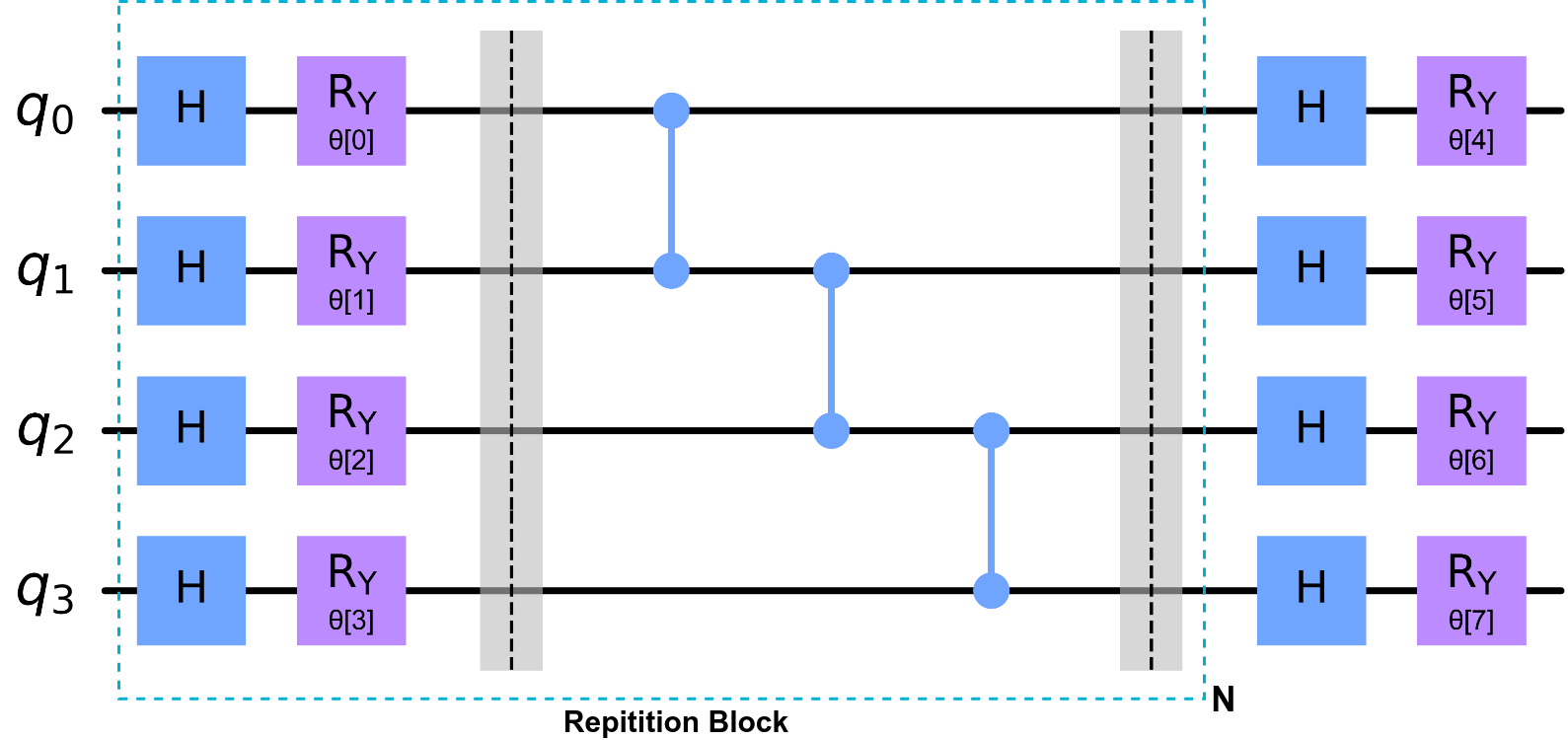}
    \caption{The TwoLocal ansatz employed in VQD algorithm, comprising of the Hadamard (H), and Y-axis rotation (RY) gates, and the Controlled-Z (CZ) gates with $N$ repetition blocks.}
\label{fig:ansatz}
\end{figure}

Since the VQD algorithm involves a sequential application of the VQE algorithm, one has to employ an ansatz that introduces the trainable parameters to the quantum circuit~\cite{vqd, vqa}. In this work, we have employed a simple hardware-efficient TwoLocal ansatz~\cite{twolocal}, see Figure~\ref{fig:ansatz}. 
The number of repetitions controls the total number of parameters of the circuit and, thereby, the overall size of the ansatz as well as the circuit depth. 
A reduced circuit depth is desirable, as it accelerates the computation and minimizes the negative impacts of quantum measurement errors, such as decoherence and noise. Furthermore, a small circuit depth enhances the scalability of quantum algorithms, making them more suitable for large-scale quantum computation. Table~\ref{tab:depth} presents the qubit requirements, the number of repetition blocks, and the circuit depth of the employed ansatz for different molecules considered in this work. It can be noted that with the compact encoding method, HMO theory can be executed with VQD algorithm on a quantum circuit with fairly small circuit depth~\cite{depth1, depth2}, hence making the process feasible and scalable even for large molecules.

\begin{table}
\caption{\label{tab:depth} The qubit requirements, the number of repetition blocks in the employed ansatz, and the circuit depth of the ansatz for quantum simulation of HMO theory of different molecules.}
\begin{ruledtabular}
    \begin{tabular}{c|ccc}

Molecule & \# & \# Repetition & Circuit \\
         & Qubits & Blocks & Depth \\
\hline  
\hline 
       C$_2$H$_4$  & 1 & 2 & 6\\
       C$_3$H$_4$  & 2 & 4 & 14\\
       C$_4$H$_4$  & 2 & 4 & 14\\
       C$_4$H$_6$  & 2 & 4 & 14\\
       C$_3$H$_4$O  & 2 & 4 & 14\\
       C$_4$H$_4$O  & 3 & 6 & 32\\
       C$_6$H$_6$  & 3 & 6 & 32\\
       C$_8$H$_{10}$ & 3 & 6 & 32\\
       C$_{60}$  & 6 & 9 & 77\\
    \end{tabular}
\end{ruledtabular}
\end{table}

\subsection{Accuracy of HMO Energy Levels}
Table~\ref{tab:table2} provides the HMO energy levels from the exact and VQD simulations of different molecules, highlighting excellent agreement of the classical exact and quantum results, even in the presence of noise.  One of the major advantages of the current encoding scheme is its versatility, that is, its ability to accommodate not only linear chains, or simple rings, but complicated structures and even hetero-atomic $\pi$-conjugated molecules. 
It can be observed that, for molecules up to \ce{C_3H_4O}, the VQD results are highly accurate across different quantum simulation settings. For these smaller molecules, the relative errors across different energy levels remain up to $10\%$, even in the realistic setting, which is impressive considering the fact that no error mitigation technique is employed in the circuits. For the molecules \ce{C_4H_4O}, \ce{C_6H_6}, and \ce{C_8H_{10}} however, the overall error becomes $20-40\%$, which can be attributed to an increase in the number of qubits and the circuit depth of the ansatz, error accumulation in the VQD algorithm, and the absence of any attempt for error mitigation. 
From Table~\ref{tab:table2}, it can be observed that the higher energy levels of larger molecules are more erroneous, a trend that can be ascribed to the VQD algorithm itself. A correction scheme for these errors is discussed in Section~\ref{sec:symvqd}.

\subsection{Molecular Orbital Construction}
In the context of HMO theory, the nature of the molecular orbitals is extensively used for the prediction of bonding and reactivity of the conjugated systems.
A major challenge for any variational algorithm is to avoid converging into a local minimum and to find the true ground state of the system.
In the case of the VQD algorithm, it is important to keep track of the converged eigenstates, as this is a sequential process where every energy level influences other measurements. Extracting the quantum state from a quantum circuit poses several challenges due to the inherent complexities of quantum mechanics and the nature of quantum information, where a general process would involve finding the probability distribution first, and then the relative phases, although this can be extremely expensive~\cite{qpe1, qpe2}. However, simulators like Qiskit's statevector\_simulator make it possible to find the exact state of a quantum circuit. For this, after convergence of the VQD algorithm to each energy level, we extracted the optimized parameters ($\{\Vec{\theta}_k\}$), and the statevector\_simulator is then employed on separate quantum circuits with similar ansatzes, set to the optimized parameters to find the eigenstates at different energy levels. Table~\ref{tab:states} presents the eigenstates from classical and quantum simulations for different molecules. The agreement between them is excellent for small molecules and remains good even for larger molecules (data not shown).

\begin{table*}
\caption{\label{tab:states} The HMO orbital energies and coefficients obtained from direct diagonalization of HMO matrix (exact) and VQD simulation with ideal quantum circuit simulator. The MO energies are in the units of resonance integral $\beta$.}
\begin{ruledtabular}
    \begin{tabular}{ccccc}
Molecule  & MO Energy & \multicolumn{2}{c}{MO Coefficients} \\ \cline{3-4}
  & & Exact  & Quantum\\
\hline  
\hline  
C$_2$H$_4$   & $-$1.0 & $[0.7071, 0.7071]^T$ & $[0.7071, 0.7070]^T$ \\
   & 1.0 & $[-0.7071, 0.7071]^T$ & $[-0.7071, 0.7071]^T$  \\
\hline  
C$_4$H$_4$ &   $-$2.0 & $[0.5, 0.5, 0.5, 0.5]^T$ & $[0.4997, 0.4999, 0.5001, 0.5]^T$\\
    & 0 & $[-0.7071, 0, 0.7071, 0]^T$ & $[-0.7068, 0.0270, 0.7063, -0.0273]^T$ \\
    & 0 & $[0, -0.7071, 0, 0.7071]^T$ & $[-0.0272, -0.7065, 0.0271, 0.7065]^T$ \\
    & 2.0 & $[-0.5, 0.5, -0.5, 0.5]^T$ & $[-0.4999, 0.5, -0.5, 0.4999]^T$ \\
\hline  
C$_4$H$_6$ &  $-$1.6180 & $[0.3717, 0.6015, 0.6015, 0.3717]^T$ & $[0.3719, 0.6015, 0.6013, 0.3716]^T$\\
    & $-$0.6180 & $[-0.6015, -0.3717, 0.3717, 0.6015]^T$ & $[-0.6013, -0.3716, 0.3719, 0.6015]^T$ \\
    & 0.6180 & $[0.6015, -0.3717, -0.3717, 0.6015]^T$ & $[0.6014, -0.3718, -0.3717, 0.6014]^T$ \\
    & 1.6180 & $[0.3717, -0.6015, 0.6015, -0.3717]^T$ & $[-0.3717, 0.6014, -0.6015, 0.3717]^T$ 
\label{table4}
\end{tabular}
\end{ruledtabular}
\end{table*}

\subsection{Simulating Large Systems Under HMO Theory with VQD}\label{sec:symvqd}
Despite its efficient performance in evaluating energy levels and state functions for the excited states of smaller systems, the VQD algorithm suffers from error accumulation, in particular, in the excited states of larger systems (Table~\ref{tab:table2}). We have dealt with this limitation of VQD algorithm by exploiting the symmetry of the systems. 
For a given Hamiltonian $H$ with eigenvalues $\{\lambda_i\}$ and the corresponding eigenvectors $\{v_i\}$ (i.e., $Hv_i=\lambda_iv_i$), the same eigenvector $v_i$ is also an eigenvector of $-H$ with eigenvalue $-\lambda_i$ (i.e., $-Hv_i=-\lambda_iv_i$).  
For a  H\"uckel matrix $H$ whose eigenvalues are given as  $E_n = [\lambda_0,\lambda_1,\lambda_2,\cdots,\lambda_n]$ where $\lambda_0<\lambda_1<\cdots<\lambda_n$, the VQD algorithm first evaluates $\lambda_0$, followed by $\lambda_1$ and so on. Therefore, the error accumulates the most for the highest eigenvalue. 
On the contrary, the eigenvalues of the H\"uckel matrix $-H$ follow the order 
$-\lambda_n<-\lambda_{n-1}<\cdots<-\lambda_1$ and VQD algorithm sequentially evaluates the states for $\lambda_n$, followed by $\lambda_{n-1}$ and so on. 
In this case, the error accumulation is the least for the $n$th state. 
This process can help us evaluate the entire eigenvalue spectrum by calculating half of the spectrum by simulating the matrix $H$ and the other half by simulating $-H$. This process exploits the symmetry of a matrix to minimize error accumulation in a VQD algorithm. We refer to it as symmetric VQD (symVQD).

\begin{figure*}[!]
 \includegraphics[width=\textwidth]{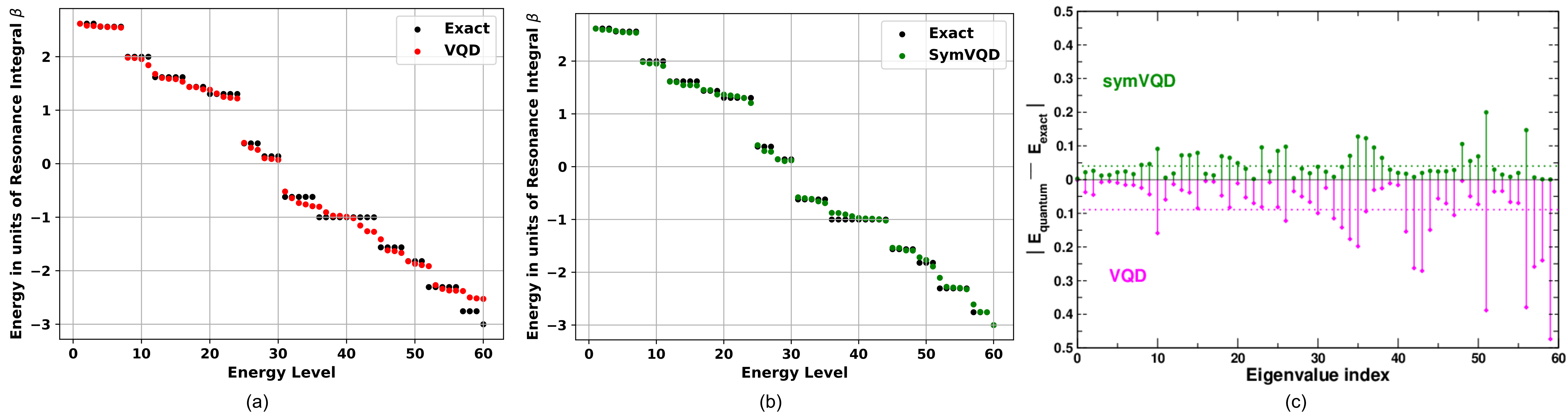}
 \caption{\label{fig:fullerene}Comparison of the exact MO energy levels of \ce{C_60} with those obtained from quantum simulation with (a) VQD and (b) symVQD algorithms on six qubits ideal quantum setting. (c) The errors in the MO energy levels of \ce{C_60} obtained from VQD (lower half of the figure) and symVQD (upper half of the figure) algorithms compared with the exact MO energy levels. The horizontal dashed lines indicate the corresponding average errors.}
\end{figure*}

The performance of our compact encoding scheme coupled with both VQD and symVQD algorithm is illustrated for $C_{60}$ fullerene, which contains 60 conjugated carbon atoms grouped in a symmetrical configuration featuring 12 pentagonal faces and 20 hexagonal faces (Figure~\ref{fig:molecules}). With the compact encoding scheme, quantum simulation of fullerene requires 6 qubits to be encoded, and the transformed Hamiltonian contains a total of 680 Pauli strings, making it a fairly large quantum system. To that end, an ansatz with 9 repetition blocks (circuit depth of 77) was employed to ensure that it is expressive enough to simulate the entire energy spectrum. 
Figure~\ref{fig:fullerene}(a), and Figure~\ref{fig:fullerene}(b) shows the energy level distribution of fullerene with both VQD and symVQD methods as compared to the exact results evaluated in an ideal quantum setting.
The symVQD algorithm is found to reproduce the exact energy level distribution of fullerene much better than the VQD algorithm (Figure~\ref{fig:fullerene}(a), Figure~\ref{fig:fullerene}(b)). This is particularly seen for the degenerate energy levels in the latter half of the distribution. The symVQD method conserves the overall structure of the distribution, showing all the degenerate levels much more clearly. On the contrary, the first half of the distribution is simulated efficiently by VQD, while the second half shows deviations from the exact results. 
The improved performance of symVQD over VQD results can be seen from the relative error $| E_i^{\rm VQD} - E_i^{\rm exact}|$ of the individual energy levels in both the algorithms, as compared to their exact values (Figure~\ref{fig:fullerene}(c)).
The average absolute error (overall energy levels) drops from 0.089 to 0.043 (in the units of $|\beta|$), around $50\%$ increased efficiency. The improvement is particularly prominent for the later energy levels, where the accumulated error of the VQD algorithm is corrected by the symVQD algorithm. This is where the symVQD has outperformed the VQD algorithm with exceptional performance at both ends of the energy spectrum.

\section{Conclusions}
This work provides a compact and efficient encoding scheme to execute H\"uckel molecular orbital theory on a quantum computer using the hybrid variational quantum deflation algorithm for excited state simulation. The capabilities of the method are demonstrated by evaluating the molecular orbital energies and eigenfunctions of simple linear molecules like \ce{C_2H_2}, \ce{C_4H_6}, and \ce{C_8H_{10}}, simple cyclic molecules like \ce{C_3H_4}, \ce{C_4H_4}, and \ce{C_6H_6}, and hetero-atomic molecules such as \ce{C_3H_4O} and \ce{C_4H_4O} within just three qubits. For the transformation of the H\"uckel Hamiltonian to qubit form, two different strategies were adopted. One involved a machine-learning-based method, while the other employed the Frobenius-inner-product to obtain the Pauli strings that contribute to the qubit Hamiltonian. The results from quantum simulations with different noise settings are in excellent agreement with the exact classical results, at least for small molecular systems. The error accumulation by the VQD algorithm in the high-energy excited states of large molecules posed a challenge. However, this was mitigated by introducing a slight modification to the VQD algorithm via the so-called symVQD method. By exploiting the symmetry of the system, this method evaluates half of the spectrum by simulating the matrix $H$, and the other half by simulating $-H$. The efficiency and scalability of the present approach have been illustrated by the quantum simulation of \ce{C_60} fullerene with a Hamiltonian containing 680 Pauli strings encoded on six qubits with a circuit depth of 77. 


The accuracy of the orbital energies and eigenfunctions produced using the HMO approach is limited to a semi-quantitative level. However, the HMO method remains a valuable tool in providing fundamental insights into the electronic structure of conjugated $\pi$-bonding systems, all while requiring little computer resources. The current work not only integrates quantum computing into the HMO theory for simple molecules but it introduces methods and strategies that are scalable and system-agnostic which allows a quantum simulation of molecules of arbitrary complexity. 
It is hoped that this work would inspire quantum simulation of similar problems, in different fields of science and technology, such as, spin chains, electronic band structure in materials, encoding/decoding in signal processing, designing control systems in robotics, etc., some of which are currently underway in our laboratory.

\section*{Acknowledgements}
This work used the Supercomputing facility of IIT Kharagpur established under the National Supercomputing Mission (NSM), Government of India, and supported by the Centre for Development of Advanced Computing (CDAC), Pune. HS acknowledges the Ministry of Education, Govt. of India, for the Prime Minister's Research Fellowship (PMRF), Ms.~Parul Dhir (PTU Jalandhar) for helpful discussions. \\

\bibliography{main.bib}

\end{document}